\documentclass[slac_one]{revtex4}
\usepackage{graphicx}
\usepackage{fancyhdr}
\begin{document}

\title{First Results from the PAMELA Space Mission} 

%


\author{M. Boezio}
\affiliation{Physics Department of University of Trieste, I-34127 Trieste, Italy}
\affiliation{INFN, Sezione di Trieste, I-34012 Trieste, Italy}

\author{O. Adriani, M. Bongi, L. Bonechi, S. Bottai, D. Fedele, P. Papini,\\ S. B. Ricciarini, P. Spillantini, E. Taddei, and E. Vannuccini}
\affiliation{Physics Department of University of Florence, I-50019 Sesto Fiorentino, Florence, Italy}
\affiliation{INFN, Sezione di Florence, I-50019 Sesto Fiorentino, Florence, Italy}

\author{G. C. Barbarino, D. Campana, R. Carbone, and G. Osteria}
\affiliation{Physics Department of University of Naples ``Federico II'',  I-80126 Naples, Italy}
\affiliation{INFN, Sezione di Naples,  I-80126 Naples, Italy}

\author{G. A. Bazilevskaya}
\affiliation{Lebedev Physical Institute,  RU-119991 Moscow, Russia}

\author{R. Bellotti}
\affiliation{Physics Department of University of Bari, I-70126 Bari, Italy}
\affiliation{INFN, Sezione di Bari, I-70126 Bari, Italy}

\author{E. A. Bogomolov}
\affiliation{Ioffe Physical Technical Institute,  RU-194021 St. Petersburg, Russia}



\author{V. Bonvicini}
\affiliation{Physics Department of University of Trieste, I-34127 Trieste, Italy}
\affiliation{INFN, Sezione di Trieste, I-34012 Trieste, Italy}

\author{S. Borisov}
\affiliation{INFN, Sezione di Roma ``Tor Vergata'', I-00133 Rome, Italy}
\affiliation{Physics Department of University of Rome ``Tor Vergata'',  I-00133 Rome, Italy}


\author{A. Bruno}
\affiliation{Physics Department of University of Bari,  I-70126 Bari, Italy}
\affiliation{INFN, Sezione di Bari, I-70126 Bari, Italy}

\author{F. Cafagna}
\affiliation{Physics Department of University of Bari,  I-70126 Bari, Italy}
\affiliation{INFN, Sezione di Bari, I-70126 Bari, Italy}



\author{P. Carlson}
\affiliation{KTH, Department of Physics, AlbaNova University Centre, SE-10691 Stockholm, Sweden}

\author{M. Casolino}
\affiliation{INFN, Sezione di Roma ``Tor Vergata'', I-00133 Rome, Italy}
\affiliation{Physics Department of University of Rome ``Tor Vergata'',  I-00133 Rome, Italy}

\author{G. Castellini}
\affiliation{ IFAC,  I-50019 Sesto Fiorentino, Florence, Italy}

\author{M. P. De Pascale}
\affiliation{INFN, Sezione di Roma ``Tor Vergata'', I-00133 Rome, Italy}
\affiliation{Physics Department of University of Rome ``Tor Vergata'',  I-00133 Rome, Italy}

\author{N. De Simone}
\affiliation{INFN, Sezione di Roma ``Tor Vergata'', I-00133 Rome, Italy}
\affiliation{Physics Department of University of Rome ``Tor Vergata'',  I-00133 Rome, Italy}

\author{\\V. Di Felice}
\affiliation{INFN, Sezione di Roma ``Tor Vergata'', I-00133 Rome, Italy}
\affiliation{Physics Department of University of Rome ``Tor Vergata'',  I-00133 Rome, Italy}


\author{A. M. Galper}
\affiliation{Moscow Engineering and Physics Institute,  RU-11540 Moscow, Russia} 

\author{L. Grishantseva}
\affiliation{Moscow Engineering and Physics Institute,  RU-11540 Moscow, Russia} 

\author{P. Hofverberg}
\affiliation{KTH, Department of Physics, AlbaNova University Centre, SE-10691 Stockholm, Sweden}

\author{G. Jerse}
\affiliation{Physics Department of University of Trieste, I-34127 Trieste, Italy}
\affiliation{INFN, Sezione di Trieste, I-34012 Trieste, Italy}

\author{A. Leonov}
\affiliation{Moscow Engineering and Physics Institute,  RU-11540 Moscow, Russia} 

\author{S. V. Koldashov}
\affiliation{Moscow Engineering and Physics Institute,  RU-11540 Moscow, Russia} 

\author{S. Y. Krutkov}
\affiliation{Ioffe Physical Technical Institute,  RU-194021 St. Petersburg, Russia}

\author{A. N. Kvashnin}
\affiliation{Lebedev Physical Institute,  RU-119991 Moscow, Russia}

\author{V. Malvezzi}
\affiliation{INFN, Sezione di Roma ``Tor Vergata'', I-00133 Rome, Italy}
\affiliation{Physics Department of University of Rome ``Tor Vergata'',  I-00133 Rome, Italy}

\author{L. Marcelli}
\affiliation{INFN, Sezione di Roma ``Tor Vergata'', I-00133 Rome, Italy}
\affiliation{Physics Department of University of Rome ``Tor Vergata'',  I-00133 Rome, Italy}

\author{W. Menn}
\affiliation{Physics Department of Universit\"{a}t Siegen, D-57068 Siegen, Germany}

\author{V. V. Mikhailov}
\affiliation{Moscow Engineering and Physics Institute,  RU-11540 Moscow, Russia} 

\author{E. Mocchiutti}
\affiliation{Physics Department of University of Trieste, I-34127 Trieste, Italy}
\affiliation{INFN, Sezione di Trieste, I-34012 Trieste, Italy}

\author{N. Nikonov}
\affiliation{INFN, Sezione di Roma ``Tor Vergata'', I-00133 Rome, Italy}
\affiliation{Physics Department of University of Rome ``Tor Vergata'',  I-00133 Rome, Italy}

\author{S. Orsi}
\affiliation{KTH, Department of Physics, AlbaNova University Centre, SE-10691 Stockholm, Sweden}



\author{M. Pearce}
\affiliation{KTH, Department of Physics, AlbaNova University Centre, SE-10691 Stockholm, Sweden}

\author{P. Picozza}
\affiliation{INFN, Sezione di Roma ``Tor Vergata'', I-00133 Rome, Italy}
\affiliation{Physics Department of University of Rome ``Tor Vergata'',  I-00133 Rome, Italy}

\author{M. Ricci}
\affiliation{INFN, Laboratori Nazionali di Frascati, Via Enrico Fermi 40, I-00044 Frascati, Italy}


\author{M. Simon}
\affiliation{Physics Department of Universit\"{a}t Siegen, D-57068 Siegen, Germany}

\author{R. Sparvoli}
\affiliation{INFN, Sezione di Roma ``Tor Vergata'', I-00133 Rome, Italy}
\affiliation{Physics Department of University of Rome ``Tor Vergata'',  I-00133 Rome, Italy}


\author{Y. I. Stozhkov}
\affiliation{Lebedev Physical Institute,  RU-119991 Moscow, Russia}


\author{A. Vacchi}
\affiliation{Physics Department of University of Trieste, I-34127 Trieste, Italy}
\affiliation{INFN, Sezione di Trieste,  I-34012 Trieste, Italy}


\author{G. Vasilyev}
\affiliation{Ioffe Physical Technical Institute, RU-194021 St. Petersburg, Russia}

\author{\\S. A. Voronov}
\affiliation{Moscow Engineering and Physics Institute,  RU-11540 Moscow, Russia}  

\author{Y. T. Yurkin}
\affiliation{Moscow Engineering and Physics Institute,  RU-11540 Moscow, Russia}  

\author{G. Zampa}
\affiliation{Physics Department of University of Trieste, I-34127 Trieste, Italy}
\affiliation{INFN, Sezione di Trieste,  I-34012 Trieste, Italy}

\author{N. Zampa}
\affiliation{Physics Department of University of Trieste, I-34127 Trieste, Italy}
\affiliation{INFN, Sezione di Trieste,  I-34012 Trieste, Italy}

\author{V. G. Zverev}
\affiliation{Moscow Engineering and Physics Institute,  RU-11540 Moscow, Russia}

\begin{abstract}
On the 15$^{th}$ of June 2006, the PAMELA satellite--borne experiment was 
launched from the Baikonur cosmodrome and it has been collecting data 
since July 2006. The apparatus comprises a time--of--flight system, a 
silicon--microstrip magnetic spectrometer, a silicon--tungsten 
electromagnetic calorimeter, an anticoincidence system, a shower tail 
counter scintillator and a neutron detector. The combination of these 
devices allows precision studies of the charged cosmic radiation to be 
conducted over a wide energy range (100 MeV -- 100's GeV) with high 
statistics. The primary scientific goal is the measurement of the 
antiproton and positron energy spectrum in order to search for exotic 
sources, such as dark matter particle annihilations. PAMELA is also 
searching for primordial antinuclei (anti--helium), and testing cosmic--ray propagation models through precise measurements of the antiparticle energy 
spectrum and precision studies of light nuclei and their isotopes. We review the status of the apparatus and present preliminary results concerning antiparticle measurements and dark--matter indirect searches.
\end{abstract}

\maketitle

\thispagestyle{fancy}

\section{ INTRODUCTION } 
Since the discovery of antimatter many efforts have been performed in order to study its abundance in the cosmic radiation, aiming at investigate the baryon symmetry in the Universe. 
Secondary positrons and antiprotons can be produced during collisions of energetic cosmic--ray particles, mainly protons, with hydrogen and helium nuclei in the interstellar medium.  
Additional sources might contribute to the local cosmic--ray antimatter abundance, like annihilation of dark--matter particles~\cite{jungman,bertone} and evaporation of primordial black holes~\cite{hawking,kiraly}.
This possibility was first suggested after the antiproton excess observed in the 1970s by the pioneering experiments of Bogomolov et al.~\cite{bogo} and Golden et al.~\cite{golden}. 
Subsequent experiments confuted these early results.
Nevertheless the idea is still alive. 
Many theoretical works support the possibility that anomalies in the spectra of rare cosmic--ray components (like antiprotons and positrons) might indicate signature of exotic processes; 
an essential requirement is that the measurements are precise enough to identify spectral structures relative to the expected secondary abundance.
No antinuclei have ever been detected in cosmic rays, which most likely excludes the presence of large domains of antimmater in the nearby region of the Universe. 
An observation of antihelium would be a significant discovery as it could indicate the presence of antimatter domains in a baryon
symmetric Universe, higher Z antinuclei (e.g. anticarbon) would be a proof of antistellar nucleosynthesis.

\section{ INSTRUMENT AND SCIENCE }
PAMELA has been mainly conceived to perform high--precision spectral measurement of antiprotons and positrons and to search for antinuclei, over a wide energy range. 

\begin{figure}[t] 
 \begin{tabular}{cc}
  \begin{minipage}{.55\hsize}
   \begin{center}
     \includegraphics[width=.7\textwidth]{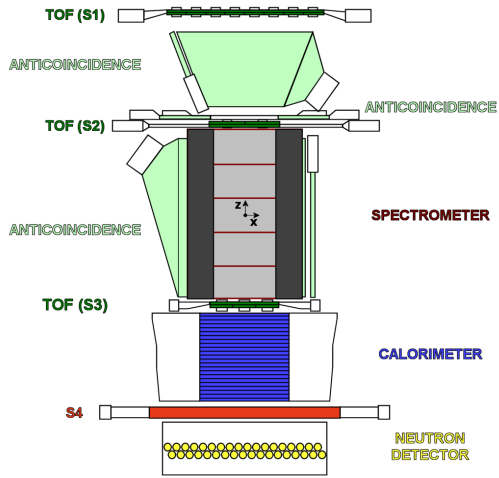}
   \end{center}
  \end{minipage}
  \begin{minipage}{.3\hsize}
   \begin{center}
     \caption{A schematic view of the PAMELA apparatus. The instruments is $\sim$1.3~m tall and has a mass of 470~kg. The average power consumption is 355~W. Magnetic field lines inside the spectrometer are oriented parallel to the y direction.\label{pamela}
}   
   \end{center}
  \end{minipage}
 \end{tabular}
\end{figure}

The PAMELA apparatus comprises the following subdetectors,
arranged as shown in Figure~\ref{pamela} (from top to bottom): 
a time--of--flight system (TOF (S1, S2,S3)); 
a magnetic spectrometer; 
an anticoincidence system (CARD, CAT, CAS); 
an electromagnetic imaging calorimeter; 
a shower tail catcher scintillator (S4) and a neutron detector. 
Planes of plastic scintillator mounted above and below the spectrometer
form the TOF system, which also provide a fast signal for triggering the data acquisition. 
The timing resolution of the TOF system allows albedo--particle identification and mass discrimination below ~1 GeV/c. 
The central part of the PAMELA apparatus is the magnetic spectrometer consisting of a 0.43~T permanent magnet and a silicon tracking system, composed of 6 planes of double--sided microstrip sensors. 
The spectrometer measures the rigidity (momentum over charge) of charged particles and the sign of the electric charge through their deflection (inverse of rigidity) in the magnetic field.  
Ionization losses are measured in the TOF 
scintillator planes, the silicon planes of the tracking system and the first silicon plane of the calorimeter allowing the absolute charge of traversing particles to be determined.  
The acceptance of the spectrometer, which also defines the
overall acceptance of the PAMELA experiment, is 21.5~cm$^{2}$sr
and the spatial resolution of the tracking system is better than 4~$\mu$m up to a zenith angle of 10$^{\circ}$, corresponding to a maximum detectable rigidity (MDR) exceeding 1~TV. 
The spectrometer is surrounded by a plastic scintillator veto shield, aiming at identify false triggers and multiparticle events generated by secondary particles produced in the apparatus. Additional information to reject multiparticle events comes from the segmentation of the TOF planes in adjacent paddles and from the tracking system.  
An electromagnetic calorimeter (16.3~X$_{0}$, 0.6~$\lambda_{0}$) mounted below the spectrometer measures the energy of incident electrons and
allows topological discrimination between electromagnetic and hadronic showers, or non--interacting particles. 
A plastic scintillator system mounted beneath the calorimeter aids in the identification of high--energy electrons and is followed by a neutron detection system for the selection of high--energy electrons which shower in the calorimeter but do not necessarily pass through the spectrometer.  
For this purpose, the calorimeter can also operate in self--trigger mode to perform an independent measurement of the lepton component up to 2~TV. 

The instrument is installed inside a pressurized container (2~mm aluminum window) attached to the Russian Resurs--DK1 Earth--observation satellite that was launched into Earth orbit by a Soyuz--U rocket on June 15$^{th}$ 2006 from the Baikonur cosmodrome in Kazakhstan. 
The satellite orbit is elliptical and quasi--polar, with an altitude varying between 350~km (north) and 610~km (south), at an inclination of 70~degrees. 
The mission is foreseen to last for at least three years.  

More technical details about the entire PAMELA instrument and launch preparations can be found in~\cite{pamelone}.

Besides the primary objective to study cosmic antimatter, the instrument setup and the flight characteristics allow many additional scientific goals to be pursued. 
Due to the high--identification capabilities of the instrument, light nuclei can be studied up to at least Z=8, as well as their isotopes. 
This provides complementary data, besides antimatter abundances, to test models for the origin and propagation of galactic cosmic rays. 
In addition, the low--cutoff orbit and long--duration mission permits to detect low--energy particles (down to 50 MeV) and to follow long--term time variations of the radiation intensity and transient phenomena. 
This allows to extend the measurements down to the solar--influence energy region,  providing unprecedented data about spectra and composition of solar energetic particles and allowing to study solar modulation of galactic cosmic rays over the minimum between solar cycles 23 and 24. 
Finally, the satellite orbit spans over a significantly large region of the Earth magnetosphere, making possible to study its effect on the incoming radiation. 
A more detailed overview of the PAMELA scientific goal can be found in~\cite{pamela2}. 

\section{ PRELIMINARY RESULTS }

To date about 600 days of data have been analyzed, corresponding to more than one billion recorded triggers. 

The main task of PAMELA is to identify antimatter components against the
most abundant cosmic--ray components. 
At high energy, main sources of background in the antimatter samples come from spillover (protons in the antiproton sample and electrons in the positron sample) and from like--charged particles (electrons in the antiproton sample and protons in the positron sample). 
Spillover background comes from the wrong determination of the charge sign
due to measured deflection uncertainty; its extent is related to the spectrometer performances and its effect is to set a limit to the maximum rigidity up to which the measurement can be extended. 
The like--charged particle background is related to the capability of the instrument to perform electron--hadron separation.

\begin{figure}[t]
  \includegraphics[width=80mm]{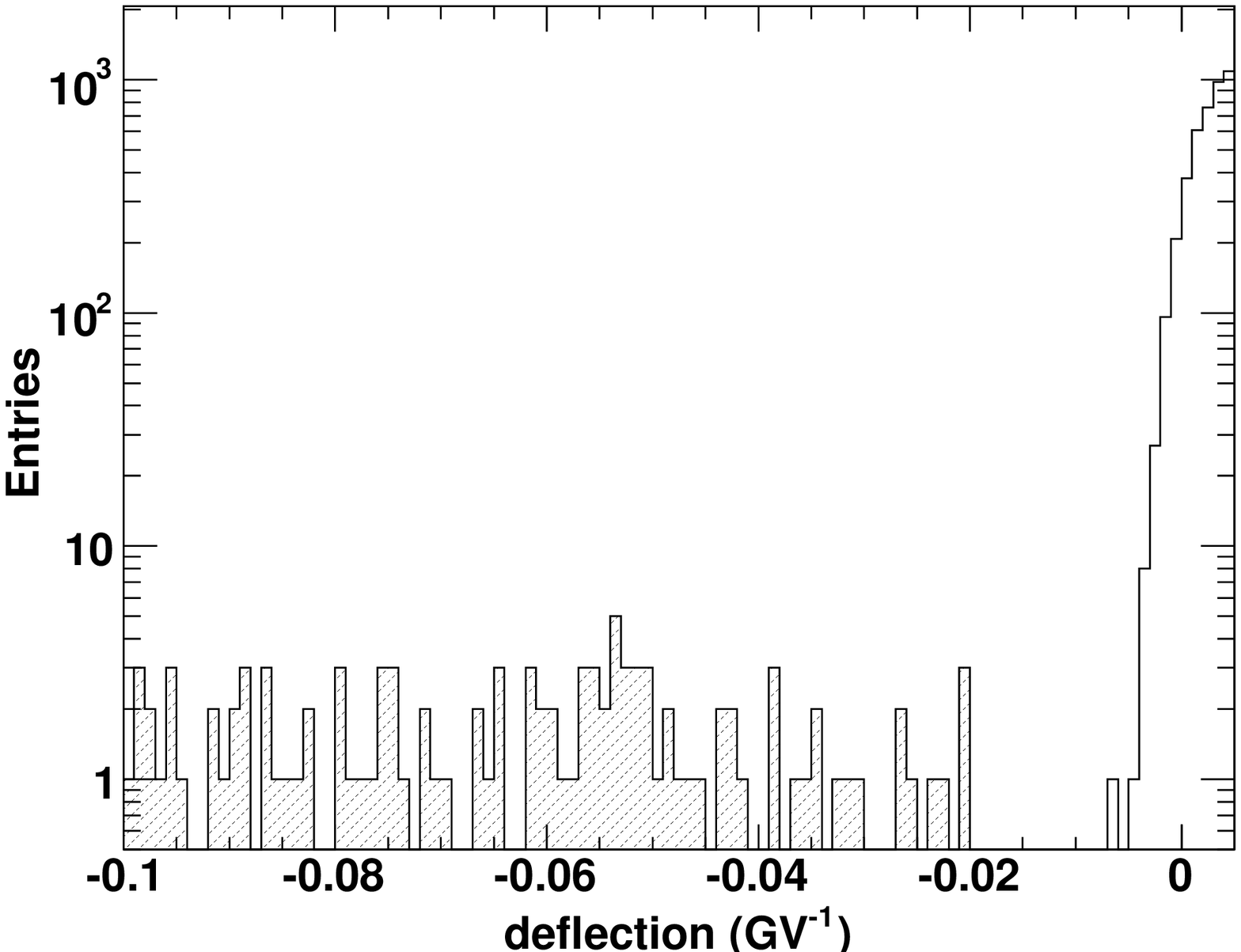}
  \includegraphics[width=80mm]{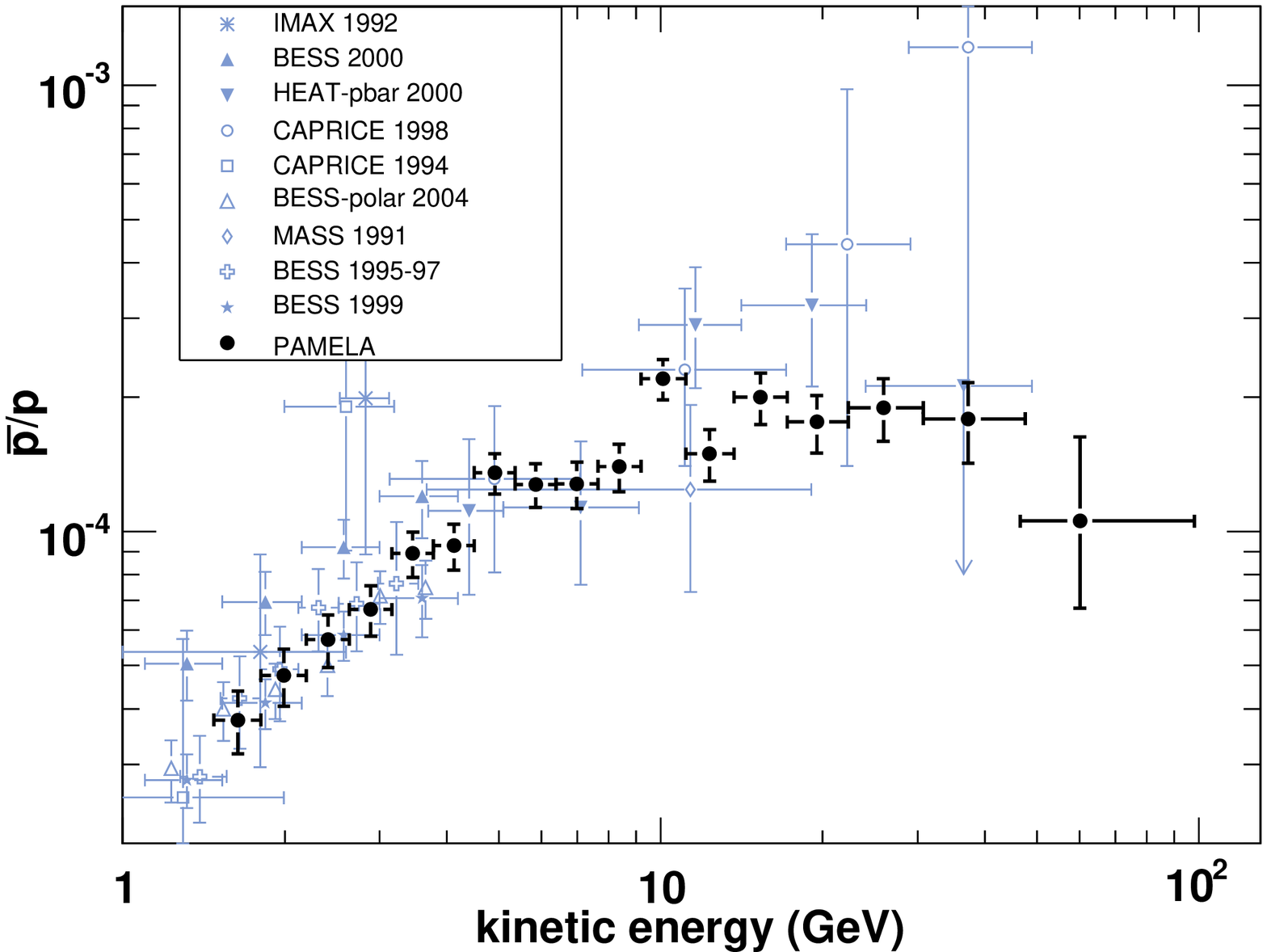} 
  \caption{Left: deflection distribution of antiprotons and protons; the sample includes events for which the evaluated deflection error is better  than 0.001177~GV$^{-1}$ (MDR~$>$~850~GV). Right: the antiproton-to-proton flux ratio obtained by PAMELA~\cite{aparticle} compared with contemporary measurements~\cite{boezio1,beach,hof,mitchell,boezio2,asaoka,hams}.\label{pbar}}
\end{figure}

Electrons in the antiproton sample can be easily rejected by applying conditions on the calorimeter shower topology, while the main source of background comes from spillover protons. 
In order to reduce the spillover background and accurately measure antiprotons up to the highest possible energy, strict selection criteria were imposed on the quality of the fitted track and the absolute value of the measured deflection was required to be 10 times larger than its estimated error. 
The deflection distribution for positively and negatively--charged down--going particles, which did not produce an electromagnetic shower in the calorimeter, is shown in Figure~\ref{pbar} (left).
The good separation between negatively--charge particles and spillover protons is evident. 

To measure the antiproton--to--proton flux ratio the different calorimeter selection efficiencies for antiprotons and protons 
were estimated. The difference is due to the momentum dependent interaction cross sections for the two particles.
These efficiencies were studied using both simulated antiprotons and protons, and proton
samples selected from the flight data. In this way, it was possible to normalize the simulated proton, and therefore the antiproton, selection efficiency.

The selected proton and antiproton samples could be contaminated by pions produced by cosmic-ray interactions with the
PAMELA payload. 
This contamination was studied using both simulated and flight data. 
At low energy, below 1~GV, negatively and positively--charged pions were identified in the flight data using the velocity measurement of the ToF system once the calorimeter rejected electrons and positrons from the sample. 
Since the majority of these pion events are produced locally in the PAMELA structure or pressure vessel it was possible to reject most of them by means of strict selection criteria on the AC scintillators and on the energy deposits in either S1 or S2.
The energy momentum spectrum of the surviving pions was measured below 1 GV and compared with the corresponding spectrum obtained from simulation. 
In the simulation, protons impinged isotropically on PAMELA from above and from the side. The protons 
were generated according to the experimental proton spectrum measured by PAMELA and 
for the pion production both GHEISHA and FLUKA generators \cite{petter,alessandro} were considered. 
The comparison between simulated and experimental pion spectrum below 1~GV resulted in a normalization factor for the simulation which accounted for all uncertainties related to the pion production and hadronic interactions. This procedure allowed to estimate the residual pion contamination on all the energy range, resulting to be less than 5\% above 2~GV decreasing to less than 1\% above 5~GV. 
It was possible to cross--check this result using flight data between 4 and 8 GV. 

Figure \ref{pbar} (right) shows the antiproton--to--proton flux ratio measured by the PAMELA experiment \cite{aparticle}
compared with other contemporary measurements. Only statistical errors are shown since the systematic uncertainty is less than a few percent of the signal, which is significantly lower than the statistical uncertainty.

The PAMELA data are in excellent agreement with recent data from other experiments,
the antiproton--to--proton flux ratio increases smoothly with energy up to about 10 GeV and then levels off. 
The data follow the trend expected from secondary
production calculations and our results
are sufficiently precise to place tight constraints on secondary production calculations and
contributions from exotic sources, e.g. dark matter particle annihilations.

PAMELA is continuously taking data and the mission is planned to continue until at
least December 2009. The increase in statistics will allow higher energies to be studied. An
analysis for low energy antiprotons (down to  100 MeV) is in progress and will be the topic
of future publications.

The analysis for positrons signal is under development and a paper is going to be submitted to international journal by the end of 2008.

\begin{acknowledgments}
We would like to acknowledge contributions and support from: Italian Space Agency
(ASI), Deutsches Zentrum f\"ur Luft-- und Raumfahrt (DLR), The Swedish National Space
Board, Swedish Research Council, Russian grants: RFBR grant 07-02-00992a and Rosobr
grant 2.2.2.2.8248.
\end{acknowledgments}

\end{document}